# NEW FACILITY FOR THE (n,2γ) REACTION INVESTIGATION AT THE DALAT REACTOR

Pham Dinh Khang[a], Vuong Huu Tan[b], Nguyen Xuan Hai[b], Nguyen Duc Tuan[b], Ho Huu Thang[b],

A.M. Sukhovoj[c], V.A. Khitrov[c]

[a]Hanoi University of Science, Vietnam

[b]Vietnam Atomic Energy Commision, Vietnam

[c]Frank Laboratory of Neutron Physics, Joint Institute for Nuclear Research, 141980 Dubna, Russia

***Abstract:*** *The summation amplitude of coincident pulses (SACP) method which is optimal solution to reduce compton scatter phenomenon and pairs phenomenon in the gamma spectra of nuclei decay gamma cascades was used. In the 1982, in comparision with original method [1], it was improved such as the interfacing techniques, and data analysis with aid of computer [2]. In order to get better, the fast/slow coincidence spectroscopy system was developed into a fast coincidence spectroscopy. It is advantageous and easy operation. The off-line measure results with radioactive source $^{60}$Co and on-line measure results with $^{35}$Cl target on the tangential channel of Dalat Research Reactor were showed the good abilities of this spectroscopy system.*

## I. The fast and slow coincidence spectroscopy system

The configuration of fast and slow coincidence spectroscopy system with changes that used to collect data which is based on the SACP method is shown in the Fig.1.

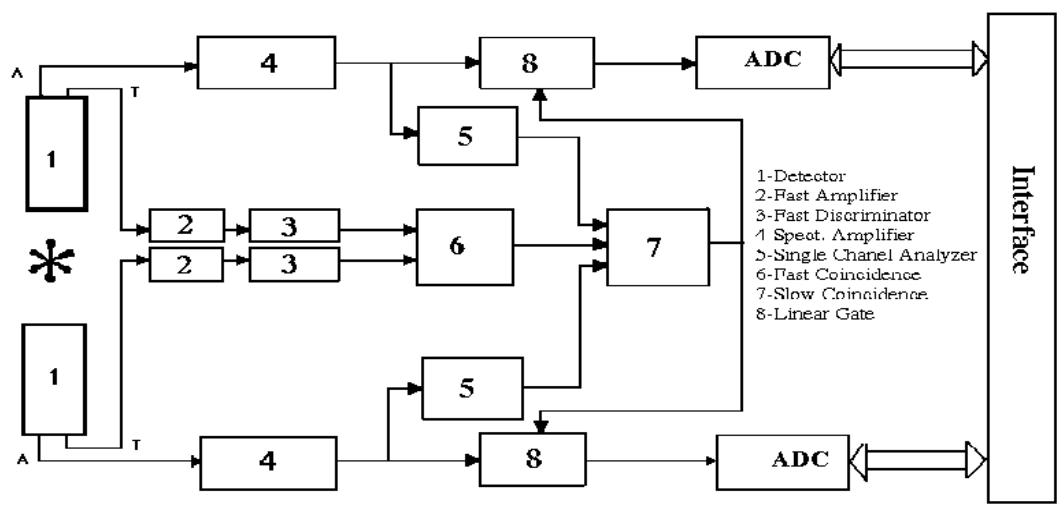

Fig. 1. Fast-Slow coincidence spectroscopy.

The function of electronic blocks which include fast filter amplifiers, constant fraction discriminators and fast coincidence is determination the two pulses from detectors which are simultaneous or not. The function of single channel analyses (SCA) is analysis energy selection of the spectroscopy. If the three output signals which one is of fast coincidence and two are from both the single channel analysis modules are simultaneous at three input of slow coincidence, the output of slow coincidence induces a pulse to open the linear gates and after that the ADCs convert input signals which are carried information of energies of gamma rays into digital values at output. Nowadays, the linear gates are combined with ADCs, and it is used in coincident mode. The range analysis of ADC is selected; it has values between low and up level of single channel analyzer. Therefore, if there is an output signal from slow coincident blocks, the ADCs convert and output data. Because of the output signal of SCA is delayed than input signal about microsecond, in order to synchronize with the record process of the pair of gamma cascade, need to use a delay line or amplifier with delay function. Unfortunately, the delay line or amplifier using affect energy resolution of spectroscopy. Consequently, if not use the delay line (or delay amplifier), slow coincidence block and SCA, that makes energy resolution effectively. In this rejection, need to solve two problems:

1. The output signal of fast amplifier (width about 20 ns) is faster than the output signal of amplifier; the interval time of difference is in order of amplifier shaping time.

2. There is an output signal of fast coincidence block but the output signal of amplifier is out of ranges of ADC's level (less than LDD or greater than LLD) therefore, only one converts the signal or both do not convert the signal. In the one ADC converting situation, the data codes are mixed. This situation is shown in the Fig. 2, the three peaks with 2346 keV and 2505 keV and 2664 keV energies in the summation spectrum of $^{60}$Co source is made.

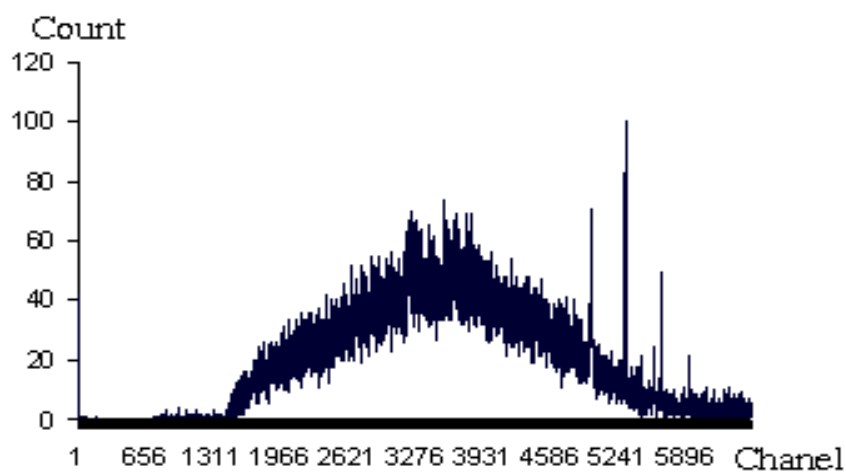

Fig. 2. There is mixing of codes in spectra $^{60}$Co.

The two problems were solved by the researchers of Vietnam Atomic Energy Commission. The spectroscopy was constructed successfully which was not used the delay line (or amplifier with delay function) and fast coincidence block and slow coincidence block and SCAs. The diagram of this spectroscopy is shown in the Fig. 3.

## II. The fast coincident spectroscopy

The following for the early output signal of coincidence:

*1. The solution for the early output signal of coincidence:* The output signals of fast coincidence block is TTL form with 500 ns width, therefore the microsecond delaying (about shaping time of the spectroscopy amplifiers) can be easily done by TTL 74123 IC.

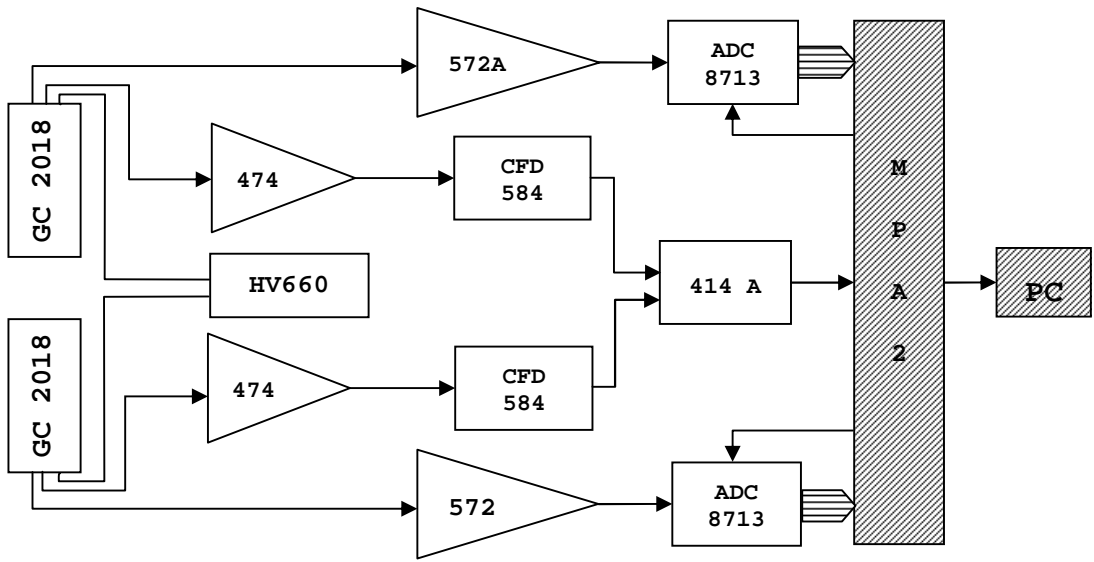

Fig. 3. The fast coincidence spectroscopy in Dalat research reactor.

*2) The solution for mixed code:*

In working of ADCs, the input signals of detectors correspond with a pair of cascade -quanta, there are three case situations:

    a. The amplitudes of signals are out of conversion range, both ADCs do not convert the input signal into output digital signal.

    b. The one amplitude of signal is out of conversion range, one ADC does not convert the input signal into output digital signal.

    c. The both ADCs convert too.

In the *a* case situation, there are not any data output from ADCs, therefore the interface does not to read and to write. In the *b* case situation, the output data is only read but not written into buffer. In the *c* case situation, after finish of reading and writing into buffer, interface induces a signal Data Accepted to both ADCs.

The control of ADC checking, reading with writing or reading without writing, can done by microprocessor. However, can do that in other way with faster and simpler:

- Because the conversions time of ADCs (ADC 8713 made by industrial CANBERRA) are the same, therefore the both Data Ready signals are induced simultaneous. Used two flip-flop circuit to make two pulses have 1 microsecond width, that are rose by rising edgy of Data Ready signals, the reading process and writing process are started by the inducing signal after Data Ready signals are multiplied. It is faster and effecter than using microprocessor.

- In order to reset ADCs, use a TTL pulse with 500 ns width, it is made from decay edge of coincident pulse after being delay 35 microsecond (include total conversion time, the reading and writing time from ADCs of interface), therefore don't care about ADCs convert or not. On the other hand, don't need to use microprocessor to reset ADCs.

By improvement, the spectroscopy was simpler, saver, more economic, easier using and enough exactitude for measurement results.

The Fig. 4 is the summation spectrum of 60Co radioactive source which was colleted by two detectors; evidently there was only peak with 1173+1332 keV energy.

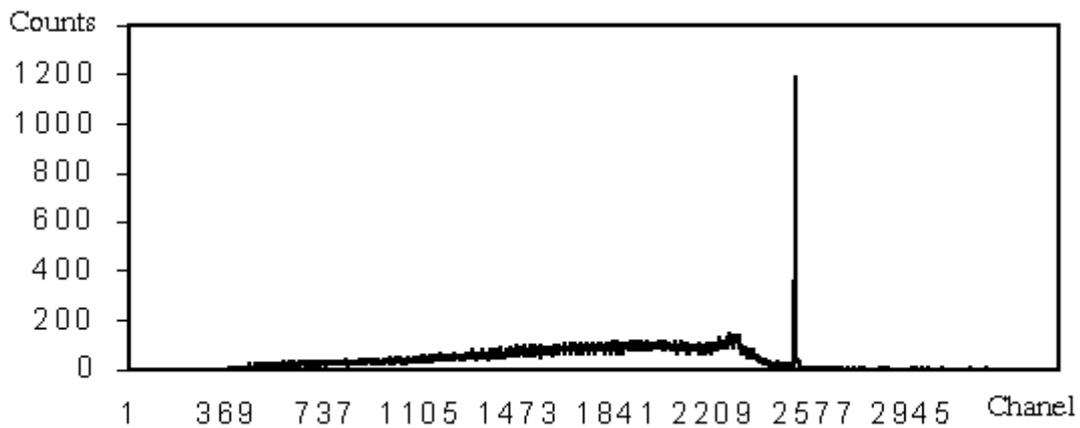

Fig. 4. The $^{60}$Co spectrum after the mixing of codes was solved.

In the Fig. 5 is the different spectrum with $^{35}$Cl target, it is measured in the on-line with some low exciting.

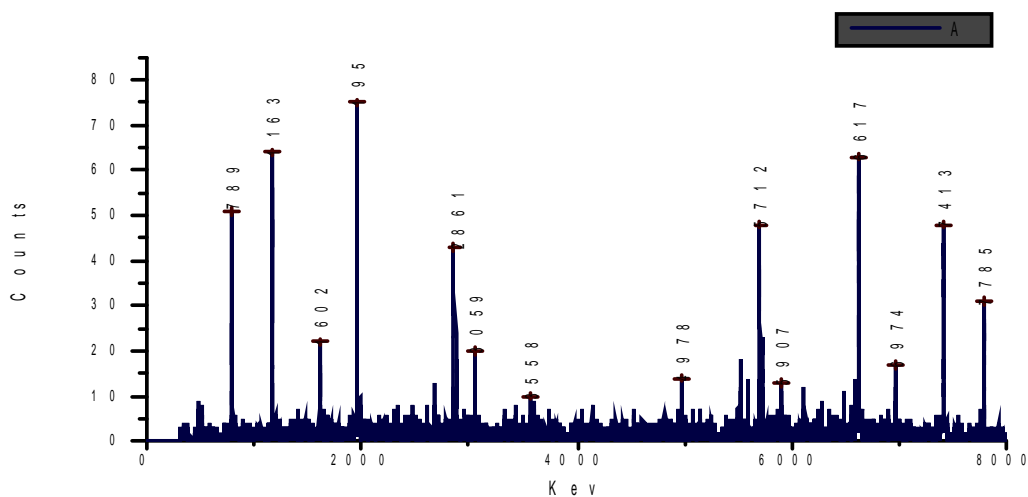

Fig. 5. The intensity distribution of two-step cascades with the total energy $E_1+E_2$=8579 keV in $^{36}$Cl (after background subtraction and correction for efficiency of registration of cascades).

## Discussion

- From the summation spectrum of $^{60}$Co radioactive source, obviousness that there is not any mixed codes in the new spectroscopy. The showing in the Fig. 4 is obeyed logical of physics process.

- The measured results with $^{35}$Cl target were compared with the results in [3], it is similar.

- The results of measurement are the same but the second configuration is simpler than fast-slow coincidence spectroscopy.

## Acknowledgements

Authors would like to thank Nuclear Research Institute and Hanoi University of Science.

The work was supported by VAEC.